# The Onset of a Globally Ice-covered State for a Land Planet

**T. Kodama[1,2], H. Genda[3], J. Leconte[2], and A. Abe-Ouchi[4,5]**

[1]Komaba Institute for Science, Graduate School of Arts and Science, The University of Tokyo, Tokyo, Japan

[2]Laboratoire d'astrophysique de Bordeaux, Université de Bordeaux, CNRS, Pessac, France

[3]Earth-Life Science Institute, Tokyo Institute of Technology, Tokyo, Japan

[4]Atmosphere and Ocean Research Institute, The University of Tokyo, Chiba, Japan

[5]Japan Agency for Marine-Earth Science and Technology, Kanagawa, Japan

Corresponding author: Takanori Kodama (koda@g.ecc.u-tokyo.ac.jp)

**Key Points:**

- The onset of a globally ice-covered state becomes lower as the area of dry region increases.
- At lower insolation, land planets are warmer than aqua planets due to fewer clouds and less snow, which leads to lower planetary albedo.
- The surface water distribution is one of the keys to determining the complete freezing limit.

## Abstract

The climates of terrestrial planets with a small amount of water on their surface, called land planets, are significantly different from the climates of planets having a large amount of surface water. Land planets have a higher runaway greenhouse threshold than aqua planets, which extends the inner edge of the habitable zone inward. Land planets also have the advantage of avoiding global freezing due to drier tropics, leading to a lower planetary albedo. In this study, we systematically investigate the complete freezing limit for various surface water distribution using a three-dimensional dynamic atmospheric model. As in a previous study, we found that a land planet climate has dry tropics that result in less snow and fewer clouds. The complete freezing limit decreases from that for aqua planets (92% $S_0$, where $S_0$ is Earth's present insolation) to that for land planets (77% $S_0$) with an increasing dry area. Values for the complete freezing limit for zonally uniform surface water distributions are consistently lower than those for meridionally uniform surface water distribution. This is because the surface water distribution in the tropics in the meridionally uniform cases causes ice-albedo feedback until a planet lapses into the complete freezing state. For a surface water distribution using the topographies of the terrestrial planets, the complete freezing limit has values near those for the meridionally uniform cases. Our results indicate that the water distribution is important for the onset of a global ice-covered state for Earth-like exoplanets.

## Plain Language Summary

Some exoplanets are thought to be Earth-like rocky planets within the habitable zone, where liquid water is stable on the planetary surface. Land planets with a small amount of surface water have the advantage of maintaining liquid water on their surface. We investigated the complete freezing limit using a three-dimensional general circulation model assuming various surface water distributions. The insolations at the complete freezing limit gradually decrease from that for a water-rich planet to that for a dry planet with increasing dry area. Our results showed that the amount of water significantly affects the initiation of a global ice-covered state for Earth-like exoplanets.

## 1 Introduction

A number of potentially habitable exoplanets located within the habitable zone of their host star have been detected. The habitable zone, which is defined as the region around a star where liquid water on a planetary surface remains stable, is the most widely used concept in exoplanetary science, as it helps narrow the search for oceans beyond Earth (e.g., Kasting et al., 1993; Kopparapu et al., 2013, 2014). In the next decade, planets within the habitable zone will be primary candidates for observation of their biosignature.

The edges of the classical habitable zone have been estimated for an aqua planet which is globally covered with ocean using a one-dimensional radiative-convective model. Recently, one can address the climate for terrestrial potential habitable planets and their habitability using three-dimensional general circulation models (GCMs). GCM studies have produced examples of climates for potential habitable planets - for example, Proxima Centauri b and the Trappist-1 planets (Turbet et al., 2016, 2018, 2020).

There are two definitions of the inner edge of the habitable zone: the runaway greenhouse limit and the moist greenhouse limit (e.g., Kasting et al., 1993; Kopparapu et al., 2013). From 1D radiative-convective equilibrium models, we know that a planet with a wet atmosphere has a limit on planetary radiation, called the Simpson-Nakajima limit, due to an atmospheric structure whereby the temperature profile asymptotically approaches the saturated vapor pressure curve for water vapor, leading to constant planetary radiation at the position where the optical depth equals unity (Ingersoll, 1969; Nakajima et al., 1992). When such a planet receives an insolation from the central star that exceeds this limit, a radiative imbalance occurs, causing higher surface temperatures (e.g., Komabayashi, 1967; Kasting, 1988; Abe & Matsui, 1988). These 1D models estimate the runaway greenhouse limit to be 282 W/m$^2$ (102% $S_0$; Goldblatt et al., 2013) and 288 W/m$^2$ (104% $S_0$; Kopparapu et al., 2013), where $S_0$ denotes the Earth's present insolation. In 3D GCM studies that predict an unsaturated downflow of water vapor in the Hadley circulation (Ishiwatari et al., 2002; Abe et al., 2011; Leconte et al., 2013a; Wolf & Toon, 2014, 2015), the runaway greenhouse limit is higher than that estimated by 1D climate models. Using LMD (Laboratoire de Météorologie Dynamique) generic GCM, Leconte et al. (2013a) estimated the runaway limit to be 110% $S_0$. Using CAM 4 (the Community Atmosphere Model version 4), Wolf and Toon (2015) showed that the climate is in thermal equilibrium up to 121% $S_0$. Way et al. (2018) showed a stable climate up to 120 $S_0$ using ROCKE-3D.

Based on the evolution of the planetary atmosphere and the luminosity of the central star, which increases with the age of the star, another limitation on the habitability of a planet arises. This is referred to as the moist greenhouse limit. When a planet with liquid water on its surface receives an insolation equal to that of Earth, the mixing ratio of water vapor is quite low above the tropopause due to the so-called cold trap mechanism. However, as the insolation received by the planet increases, water vapor can extend to a higher altitude where it is photodissociated into hydrogen. Therefore, rapid escape of hydrogen to space occurs because of the EUV radiation from the central star (Hunten, 1973; Walker, 1977). When the water vapor mixing ratio in the upper atmosphere is above $\sim 3 \times 10^{-3}$, the planet loses an amount of water equal to that in the current Earth's ocean in 4.5 billion years (Kasting et al., 1993). The state in which this water vapor mixing ratio in the upper atmosphere is achieved is called the moist greenhouse limit.

Beyond the outer edge of the habitable zone, planets are unable to maintain liquid water on their surface, leading to a so-called snowball state, in which the planetary surface is covered entirely by snow or ice. Most studies on the outer boundary of the habitable zone focus on the amount of $CO_2$ in the atmosphere, which can lead to warming of the planetary surface due to the greenhouse effect. However, the greenhouse effect due to $CO_2$ has a limitation because it also has an effect of Rayleigh scattering as cooling. When the greenhouse effect due to $CO_2$ reaches its maximum, the insolation required to keep the planet habitable is called the maximum greenhouse effect limit. Based on calculations using a 1D radiative-convective model, this limit is 0.343 $S_0$ for a $CO_2$ partial pressure of ~8 bar (Kopparapu et al., 2013). Some studies have suggested that the formation of $CO_2$ clouds, rather than the maximum greenhouse effect of $CO_2$, determines the boundary of the habitable zone. However, the effect depends on the opacity, size, and distribution of clouds, which are very difficult to predict, even on the Earth.

From theoretical studies of terrestrial exoplanet formation, such planets tend to have a larger amount of water than is present on the Earth (Raymond et al., 2004, 2007; Tian & Ida, 2015; O'Brien et al., 2018; Ikoma et al., 2018) due to the transportation of small water-rich bodies to the inner region beyond the snow line. Considering the protoplanetary disk, where water can be created via oxidation of atmospheric hydrogen caused by oxidizing minerals from planetesimals or the magma ocean, there are widely varying ranges for the amount of water that planets can have in the habitable zone (Kimura & Ikoma, 2020). Thus, for potentially habitable exoplanets, we should expect a variation in the amount of water on their surface.

While most studies on the climate of potentially habitable planets focus on aqua planets, that is, planets with a large amount of liquid water on their surface, GCMs have been used to investigate the climate of land planets - planets having a relatively small amount of water on their surface (Abe et al., 2005, 2011; Abbot et al., 2012; Leconte et al., 2013; Kodama et al., 2018, 2019; Way & Del Genio, 2020). The most important climatic characteristic of a land planet is its much drier atmosphere compared to that of an aqua planet. Whereas an aqua planet has a globally connected ocean on its surface, the liquid water on a land planet with low obliquity is localized around the polar regions. Owing to its drier atmosphere, the climate of a land planet results in a wider habitable zone than is the case for an aqua planet.

Kodama et al. (2019) focused on estimating the runaway greenhouse limit of the habitable zone for a land planet and found that it strongly depends on the distribution of surface water. The threshold of the runaway greenhouse effect increases with decreasing surface water area, from 130% $S_0$ to 180% $S_0$. This results from an increase in the limit on planetary radiation due to the distribution of water vapor in the atmosphere. Kodama et al. (2018, 2019) assumed several sets of configurations for the surface water distribution - zonally and meridionally uniform surface water distributions - and showed the varying insolation at which the runaway greenhouse occurs, corresponding to the inner edge of the habitable zone for land planets (see Fig.7 in Kodama et al., 2019).

Even within the habitable zone, liquid water on the planetary surface would freeze when the runaway glaciation occurs. With regard to Earth's history, one thought that there were at least two global-scale glaciations, one is the Marinoan glaciation at ~ 635 million years ago and the other is the Sturtian glaciation at ~ 715 million years ago, the so-called "Snowball" Earth hypothesis (e.g., Kirschvink, 1992; Hoffman et al. 1998). Many previous studies investigated the initiation of the Snowball Earth state as a function of reduced solar insolation and/or $CO_2$ concentration in the atmosphere, focusing on the runaway ice-albedo feedback. Yang et al. (2012a, b) pointed out the importance of the ice-albedo feedback and the water vapor feedback on the initiation of the Snowball Earth state. Yang et al. (2012c) also found that the runaway glaciation occurs in CCSM4 (The NCAR Community Climate System Model version 4) at an 8-9% reduced solar insolation with 286 ppmv $CO_2$ concentration or at a 6% reduced solar insolation with 70-100 ppmv $CO_2$ concentration. Voigt and Maroyzke (2010) also showed the required condition for runaway glaciation for a 6-9% reduction in solar radiation with 286 ppm $CO_2$. Additionally, the importance of the treatment for sea ice/snow albedo parameterization has been discussed with respect to the Snowball Earth bifurcation (e.g., Pierrehumbert et al. 2011; Abbot et al. 2011; Voigt and Abbot 2012; Yang et al. 2012a; 2012b).

It should be noted, however, that these studies assumed paleo-Earth conditions and thus dealt with Earth-like, water-rich aqua planets. A land planet has the advantage of avoiding global freezing. Abe et al. (2011) investigated the limit for the complete freezing of liquid water on a land planet using a GCM and found that it was ~77% $S_0$, compared to ~90% $S_0$ for an aqua planet. As for the climate of a land planet, its tropical region is much drier than that of an aqua planet, leading to a lower planetary albedo caused by a small cloud fraction and snowfall amount.

Land planets are thought to be good candidates for habitable planets because of their much wider habitable zone. Abe et al. (2011) estimated the runaway greenhouse and freezing limits, and determined the habitable zone for a typical land planet. Kodama et al. (2018, 2019) investigated the threshold of the runaway greenhouse effect for a land planet considering various surface water distributions and concluded that the runaway greenhouse limit strongly depends on the surface water distribution. The complete freezing limit must also be affected by the surface water distribution because it governs the distribution of water vapor, which strongly affects regions where rainfall and snowfall occur. In this study, we use the same configuration as Kodama et al. (2018, 2019) to investigate the complete freezing limit for various surface water distributions.

In section 2, we describe our GCM and the surface water distribution assumed in the study. In section 3, we show typical results for the complete freezing limit for cases of zonally and meridionally uniform surface water distributions. We also show results for a surface water distribution determined by terrestrial planetary topographies. In section 4, we describe our simulation for the complete freezing limit as a function of the initial land fraction and discuss the effects of the amount of $CO_2$ on the climate and exoplanets. Finally, in section 5, we summarize our findings.

## 2 Methods

We used 3D GCM, CCSR/NIES AGCM 5.4g (Center for Climate System Research/National Institute for Environmental Studies, Atmospheric General Circulation Model 5.4g) to investigate the climate for a land planet. This is the same model that was used by Abe et al. (2011) and Kodama et al. (2018, 2019). Below, we briefly describe the essential parts of GCM used in this study.

## 2.1 GCM description

The GCM used in our study was originally developed for studying Earth's climate (Numaguti, 1999). It was also applied to the paleo (Abe-Ouchi et al., 2013) and planetary climate (Abe et al., 2011; Kodama et al., 2018, 2019). The model solves the global primitive equations for dynamical processes coupled with a radiative transfer scheme. For the horizontal direction, the scheme for large-scale atmospheric dynamics uses the spectral transform method. For the vertical direction, it uses the sigma-coordinate with non-uniform spacing for its grid discretization (Arakawa & Suarez, 1983). The resolution of the GCM is T21L20, which corresponds to approximately 5.6º in longitude and latitude grids, with 20 vertical layers. In a part of the radiative transfer, the model uses a two-stream radiative transfer method with a k-distribution scheme with 18 bins and several subchannels, ranging from 50 to 50,000 $cm^{-1}$. The total number of bins is 37 (Nakajima & Tanaka, 1986).

To account for the process of cloud condensation in the atmosphere, the scheme of prognostic cloud water is applied based on the scheme results from Le Treut & Li (1991) for the large-scale condensation process. For precipitation, a simplified Arakawa-Schubert scheme is adopted for cumulus precipitation (Arakawa & Schubert, 1974). Both cumulus clouds and large-scale condensation are taken into account for the radiative transfer calculation. Whether precipitation on the surface is rain or snow is determined by whether the temperature in the lower atmospheric layer is below or above the freezing point of water. We treat soil moisture using the bucket model (Manabe, 1969). The capacity of the bucket in each grid cell is 1000 m to avoid surface runoff. On the surface, we estimate the potential evaporation using a bulk formula, which allows us to estimate evaporation efficiency. In our GCM, we set a critical soil moisture to be 10 cm, which determines wetness. When the soil moisture in a grid cell exceeds 10 cm, the grid cell is considered to be completely wet. We assume a typical surface albedo for a desert (0.3) as the surface albedo except for snow-covered region. Even though a region would have a large amount of water, we do not take account of the effect of wetness of soil on the surface albedo. The surface albedo for a snow-covered region is assumed to be proportional to the square root of the thickness of the snow. The Bond albedo for ice and snow increases from 0.5 to 0.75 with decreasing surface temperature from 273.15 to 258.15 K. The snowpack is considered as a single reservoir and is prognostically determined by the balance between snowfall and snowmelt. When the surface skin temperature is above the melting temperature, snowmelt occurs.

We removed the effect of vegetation and heat transport by the ocean from the simulation for Earth in order to build an idealized water planet. These settings are the same as those in Abe et al. (2011) and Kodama et al. (2018, 2019). We assume the $CO_2$ concentration to be 345 ppm, with a fixed 1 bar $N_2$ atmosphere. The planetary size, gravity, and orbital period are the same as the values for Earth. The planetary obliquity and eccentricity are set to zero to avoid seasonal changes in insolation.

## 2.2 Surface water distributions

We assume three different surface water distributions: a zonally uniform water distribution, a meridionally uniform water distribution, and a surface water distribution determined by specific planetary topographies, as in Kodama et al. (2018, 2019). For the zonally uniform water distribution, we use the concept of a water flow limit as defined in Kodama et al. (2018). The water flow limit is the latitude that sets the boundary between the water pool region, which is always wet, and the dry region. The planetary surface is totally wet at latitudes above the water flow limit. For the meridionally uniform water distribution, liquid water is present on the surface from the North pole to the South pole. This case is divided into two groups: the meridionally equally dispersed case and the meridionally concentrated case. For the surface water distribution resulting from the planetary topographies, we evaluate the topographies for terrestrial planets, which include Earth, Mars and Venus, using spherical harmonics, based on Wieczorek (2007) and Hirt et al. (2012). Using these topographies, we create several surface water distributions with different water amounts on the surface. Figure 1 shows examples of the surface water distributions considered in this study (for more details, see Kodama et al., 2019).

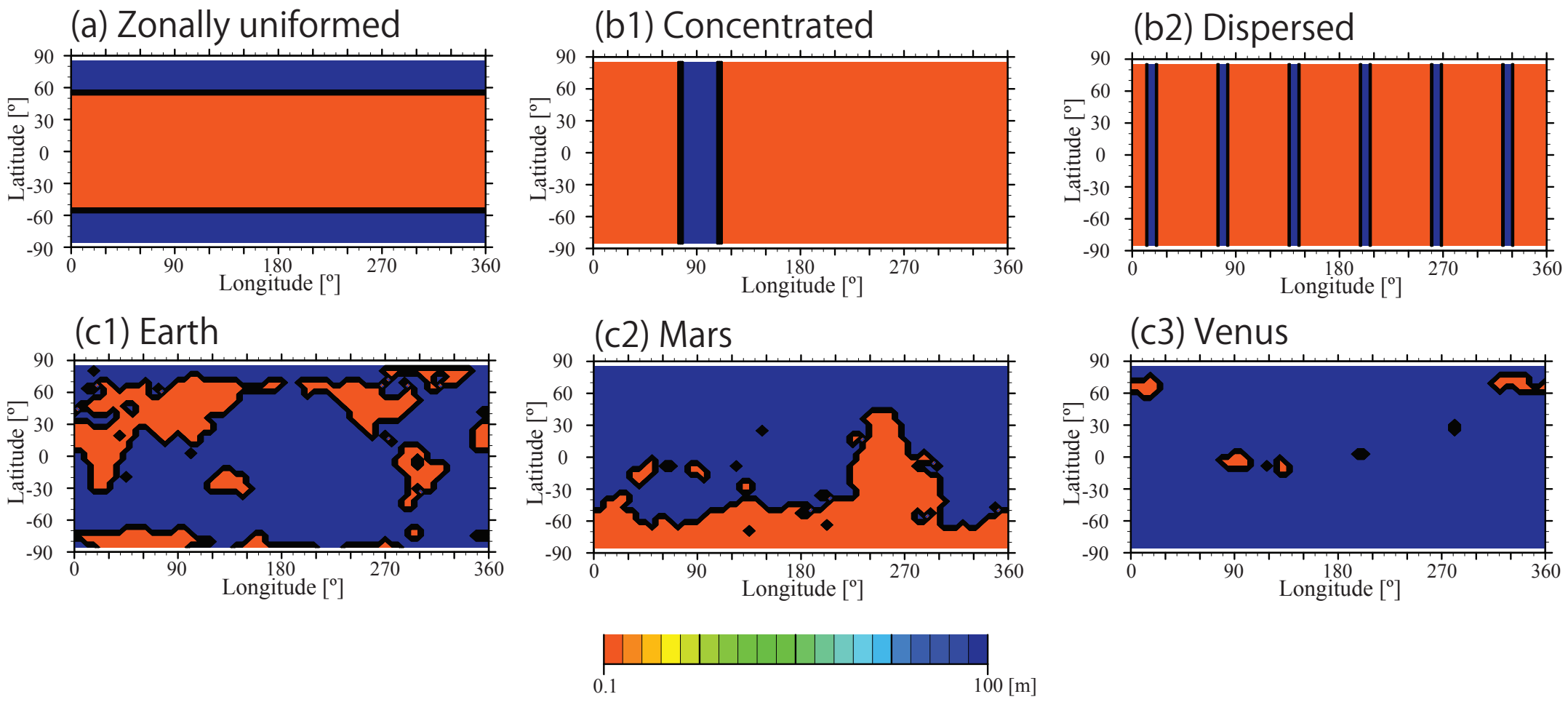

Figure 1. Example initial distributions of surface water in cases for (a) a zonally uniform case, (b) meridionally uniform cases, and (c) the cases with terrestrial planetary topographies. In the cases (b), the meridionally concentrated case (b1) and the meridionally equally-dispersed case (b2) are shown. For the cases (c), the topographies for the Earth (c1), Mars (c2), and Venus (c3) are assumed with an equivalent water amount of the present Earth. Colors represent the depth of water.

## 2.3 Numerical simulation procedure

We created the initial conditions by performing GCM calculations for an aqua planet with the insolation of the present Earth for 10 years to produce a stable climate. The initial condition is that of Kodama et al. (2018, 2019). We replaced only the surface water distribution from the initial one to that assumed for each case. Then, we ran GCM simulation with assumed the surface water distribution with 100% $S_0$ again to get stable climate for each case. After producing a stable climate, we reset the surface water distribution to consider fixed surface water distribution. We put an assumed surface water distribution with 100m depth in the water pool region. For any region out of the assumed surface water distribution where the depth of the water exceeded 10 cm, the depth is set to 10 cm. The depth in the rest region is unchanged. We decreased the insolation received by the planet in decrements of 1% $S_0$ from the atmospheric condition one before with resetting surface water distribution until the planetary climate reached the complete freezing limit. Each simulation was for 50 years. For the complete freezing limit for a land planet, we followed Abe et al. (2011), who defined it as the state in which the entire surface of the planet is permanently covered by ice or snow.

## 3 Freezing limit for a land planet

### 3.1 Zonally uniform surface water distribution cases

We investigated the climate of a land planet with a zonally uniform surface water distribution using the concept of the water flow limit being the lowest latitude at which surface water is present. Because our GCM has 32 grid cells in latitude, we created 15 cases of initial surface water distributions, symmetrical to the equator. Figure 2 shows the zonally averaged surface temperature and snow amount as a function of insolation for two water flow limit cases, 0º and 58.1º, respectively. For the 0º water flow limit, we stopped the simulation below 85% $S_0$. We can compare between behaviors of the climate on planets with different water flow limit when the insolation decreases. A planet totally covered with liquid water, which is the case for a 0º water flow limit, enters the complete freezing state at 92% $S_0$, similar to the value for an aqua planet (90% $S_0$) reported by Abe et al. (2011) (see Figs. 2a and 2c). For a planet with a 58.1º water flow limit, the surface temperature is much warmer and there is less snow than in the aqua planet case reported by Abe et al. (2011) (Figs. 1b and 1d). We found that the complete freezing limit for a planet with a 58.1º water flow limit is 77% $S_0$, comparable to the complete freezing limit for a land planet (77% $S_0$) reported by Abe et al. (2011).

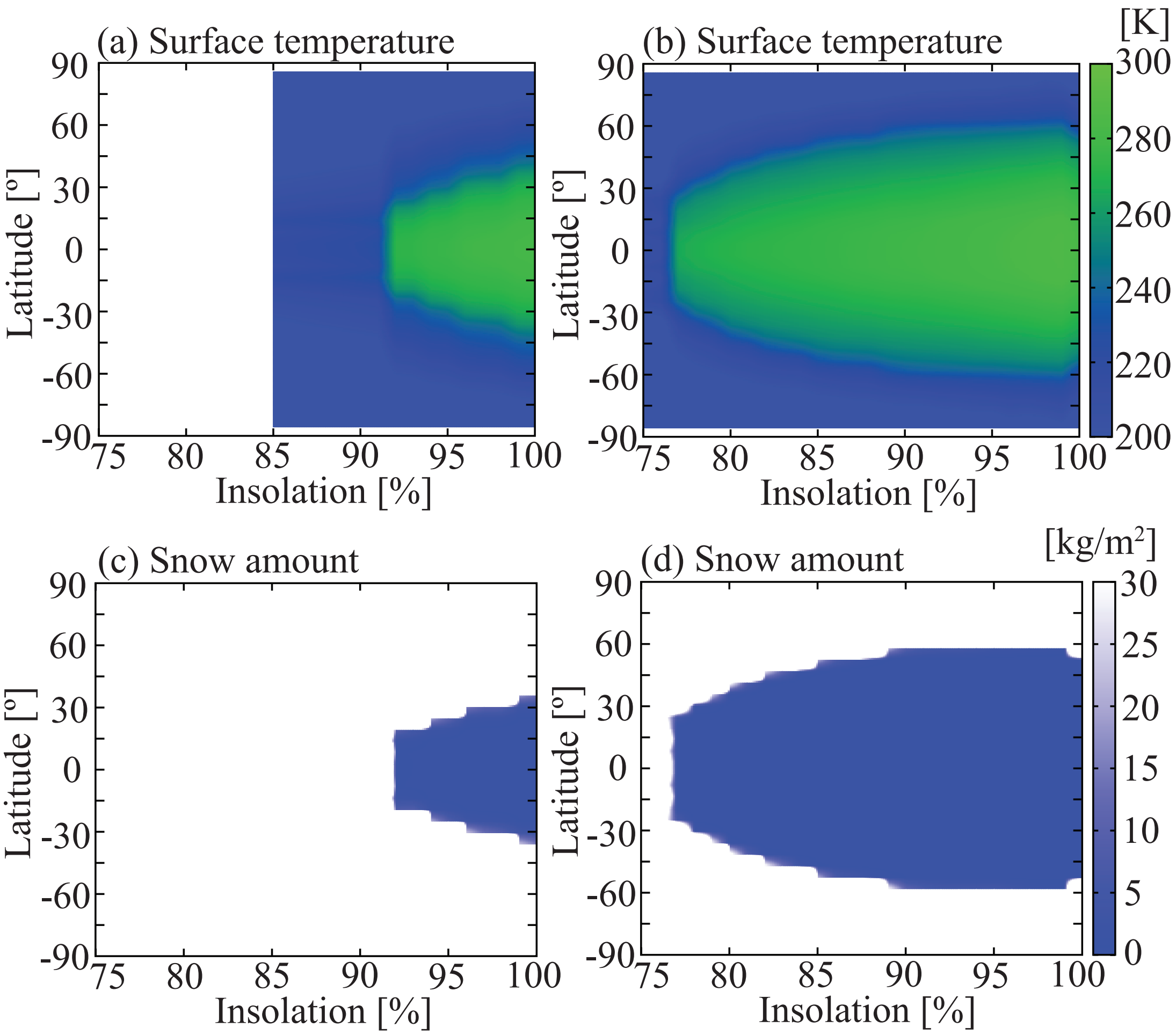

Figure 2. The zonally averaged surface temperature (upper row) and the snow amount (lower row) for cases of 0° (left) and 58.1° (right) in the water flow limit as a function of the insolation.

To understand these climates, the cloud fraction, the cloud radiative forcing, and planetary and surface albedo for the typical cases shown in Fig. 2 at the complete freezing limit are shown in Fig. 3. The difference between the freezing limits for an aqua planet and a land planet is due to the cloud and snow coverage, leading to a higher albedo, which was not shown in the figures in Abe et al. (2011) although they pointed it out. Figures 3a and 3d compare the cloud fraction for both cases at 92% $S_0$, the complete freezing limit for an aqua planet. A planet with a 0° water flow limit has more widely distributed cloud in the tropical region than a planet with a 58.1° water flow limit, as the atmosphere in the tropics for a land planet is drier due to the surface water distribution. For the cloud radiative forcing, we show the zonally averaged cloud longwave radiative forcing $\mathrm{CRF_{longwave}} = \mathrm{LW_{TOA}^{clear\ sky}} - \mathrm{LW_{TOA}^{all\ sky}}$, shortwave radiative forcing $\mathrm{CRF_{shortwave}} = \mathrm{SW_{TOA}^{all\ sky}} - \mathrm{SW_{TOA}^{clear\ sky}}$, and net radiative forcing (longwave plus shortwave). In general, clouds can cool the climate by reflecting the insolation and can warm the climate by reducing the planetary radiation. We can evaluate whether clouds cool or warm the climate from the magnitude of the clouds radiative forcing. A planet with a 58.1º water flow limit has smaller cloud radiative forcings, comparing with that for a planet with a 0º water flow limit because its atmosphere is drier and the cloud amount is smaller. Figure 3 (c) and (f) show the planetary albedo for all sky $\alpha_{\mathrm{all\ sky}} = \mathrm{SW}_{\uparrow,\mathrm{TOA}}^{\mathrm{all\ sky}}/\mathrm{SW}_{\downarrow,\mathrm{TOA}}$ and clear sky $\alpha_{\mathrm{clear\ sky}} = \mathrm{SW}_{\uparrow,\mathrm{TOA}}^{\mathrm{clear\ sky}}/\mathrm{SW}_{\downarrow,\mathrm{TOA}}$, the surface albedo, and the contribution of clouds to the planetary albedo $\alpha_{\mathrm{cloud}} = \alpha_{\mathrm{all\ sky}} - \alpha_{\mathrm{clear\ sky}}$. For a case with 0º water flow limit, the planetary albedo is much higher than in the case of a 58.1º water flow limit due to the wider snow distribution and greater contribution from clouds. The relatively low albedo for a planet with a 58.1º water flow limit thus makes the climate warmer. To analyze the greenhouse effect of water vapor, we need to disentangle the radiative effects of clouds and water vapor. The effective surface emissivity $\varepsilon$, defined as the ratio of the outgoing all sky radiation on long wavelengths at the top of the atmosphere to the upward longwave radiation from the planetary surface, can be used to denote the magnitude of the greenhouse effect as $G = 1 - \varepsilon$. For a planet with a water flow limit of 0º, the total effective surface emissivity is 0.802, versus 0.788 for the 58.1º water flow limit case. On a planet with a 58.1º water flow limit, the contribution of clouds to the total effective surface emissivity is $-3.475 \times 10^{-5}$, as compared to $-3.187 \times 10^{-2}$ for the case with a 0º water flow limit, due to the cloudless atmosphere in the 58.1º water flow limit case. Thus, the greenhouse effect in the 58.1º case is a bit larger than that in the case of the 0º water flow limit because of the higher surface temperature and the smaller radiation effect of clouds.

These are typical climatic features of an aqua planet and a land planet, respectively. Accordingly, we confirm that a planet with a 58.1º water flow limit acts as a land planet as the insolation decreases. Even land planets that have a strong resistance to decreases in surface temperature can lapse into the freezing state at some point. As shown in Fig. 3g, the cloud distribution is similar to that for an aqua planet at the complete freezing limit, but the net cloud radiative forcing is positive because of a lack of low clouds in the tropics. The total effective surface emissivity is 0.846, with a $-3.880 \times 10^{-3}$ of the contribution of clouds because of the small amount of water vapor in the atmosphere. Thus, the snow distribution causes a higher planetary albedo and a lower surface temperature, leading to complete freezing. These comparisons of aqua planet and land planet climates clearly show that a planet with less surface water distribution has a much warmer surface temperature due to the lower planetary albedo caused by the snow and cloud distributions.

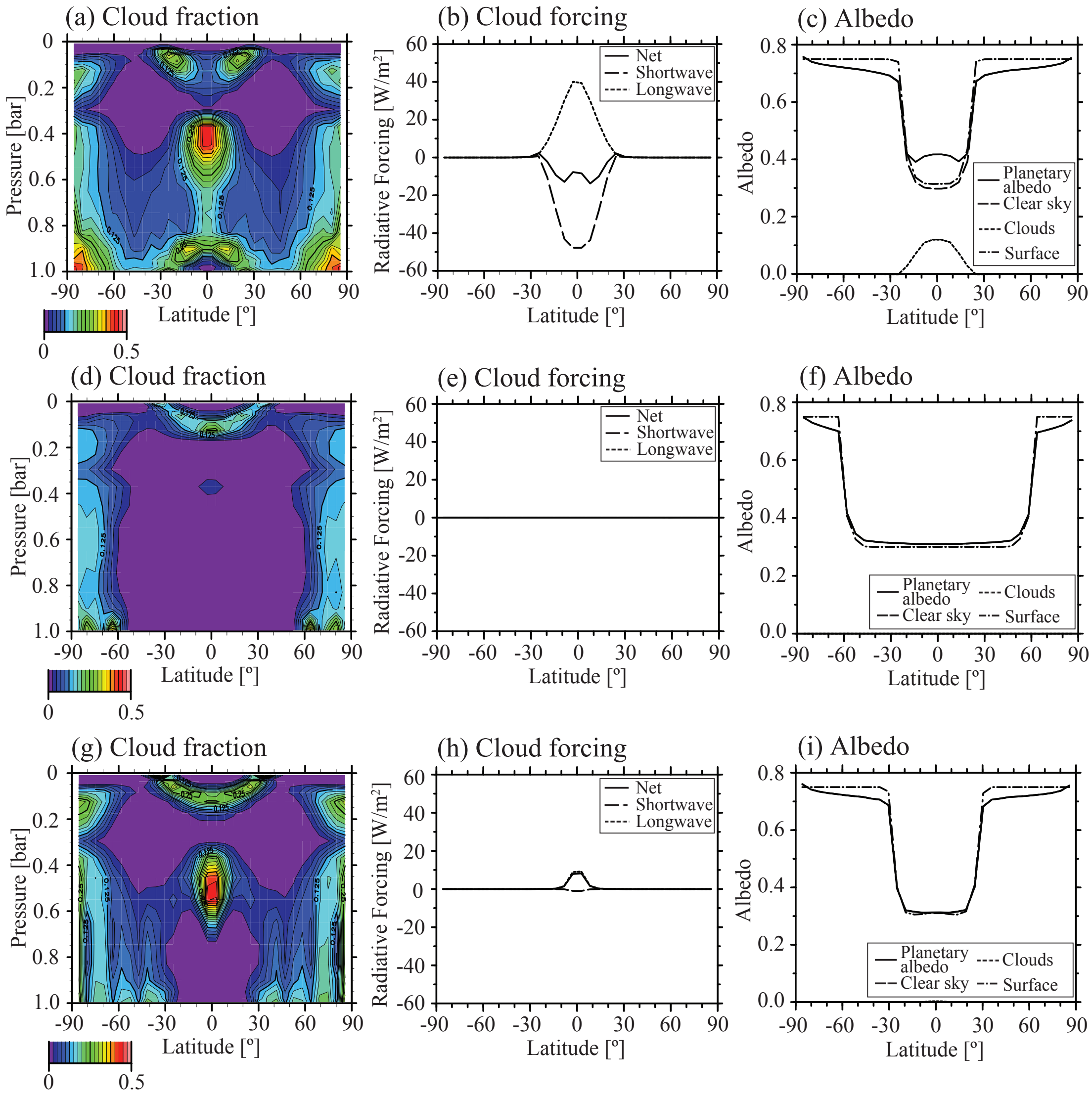

Figure 3. The zonally averaged cloud fraction (left column), the zonally averaged cloud radiative forcing (center column), and the zonally averaged albedos (right column) for 0º and 58.1º of the water flow limits with different insolation. For the cloud radiative forcing, the solid line shows the net cloud radiative forcing, the dashed line shows the cloud shortwave radiative forcing, and the dotted line shows the cloud longwave radiative forcing. For the albedos, it includes the planetary albedo for all sky and clear sky, the surface albedo, and the contribution of clouds to the planetary albedo. Upper row shows a case for 0º of the water flow limit at 92% $S_0$, which corresponds to the complete freezing limit. Middle row shows a case for 58.1º of the water flow limit at 92% $S_0$ for comparison with the case for 0º of the water flow limit case. Bottom row shows a case for 58.1º of the water flow limit at 77% $S_0$.

### 3.2 Meridional surface water distribution cases

There were also climate differences between meridional surface water distributions. For the runaway greenhouse limit, Kodama et al. (2019) showed that the meridional surface water distribution affects the climate because it controls the distribution of water vapor in the atmosphere. Thus, it is important to determine the complete freezing limit via the cloud and snow distributions. Following Kodama et al. (2019), we consider two sets of meridional surface water distributions: the meridionally concentrated case and the meridionally equally-dispersed case (b1 and b2 in Fig. 1). The water pool regions in these cases have the same area, but the water distributions are different. For example, in the meridionally concentrated case, when we set six of the 64 grid cells as water pool regions on a planetary surface, which are placed next to each other. In contrast, in the meridionally equally-dispersed case, the water pool regions are separated and equally spaced. We investigated the climate with various water pool region patterns, decreasing the insolation in order to evaluate the complete freezing limit.

Figures 4a-c show the meridionally averaged cloud fraction in the tropics as well as the cloud radiative forcing and the planetary and surface albedos at the complete freezing limit in a concentrated case for 52 grid cells of the water pool region. In this case, complete freezing occurs at 90% $S_0$, similar to the value of the complete freezing for an aqua planet as reported by Abe et al. (2011). A region with a smaller cloud fraction (Fig. 4a) appears due to a dry area outside the water pool region. Figures 4d-f show the same set of variables as Figs. 4a-c, but the number of grid cells in the water pool region is set to six, at the same insolation. These cells are set around a longitude of 90º. This planet has a much smaller cloud fraction (Fig. 4d) and weaker cloud radiative forcing (Fig. 4e) than the case with 52 grid cells in the water pool region because of the surface water distribution, leading to a lower albedo (Fig. 4f) and a higher surface temperature. Thus, for a meridional surface water distribution, a planet having a wider dry region on its surface behaves like a land planet, with a strong resistance to complete freezing due to the cloud and snow distributions. This result is consistent with Abe et al. (2011) and with the results obtained in Section 3.1.

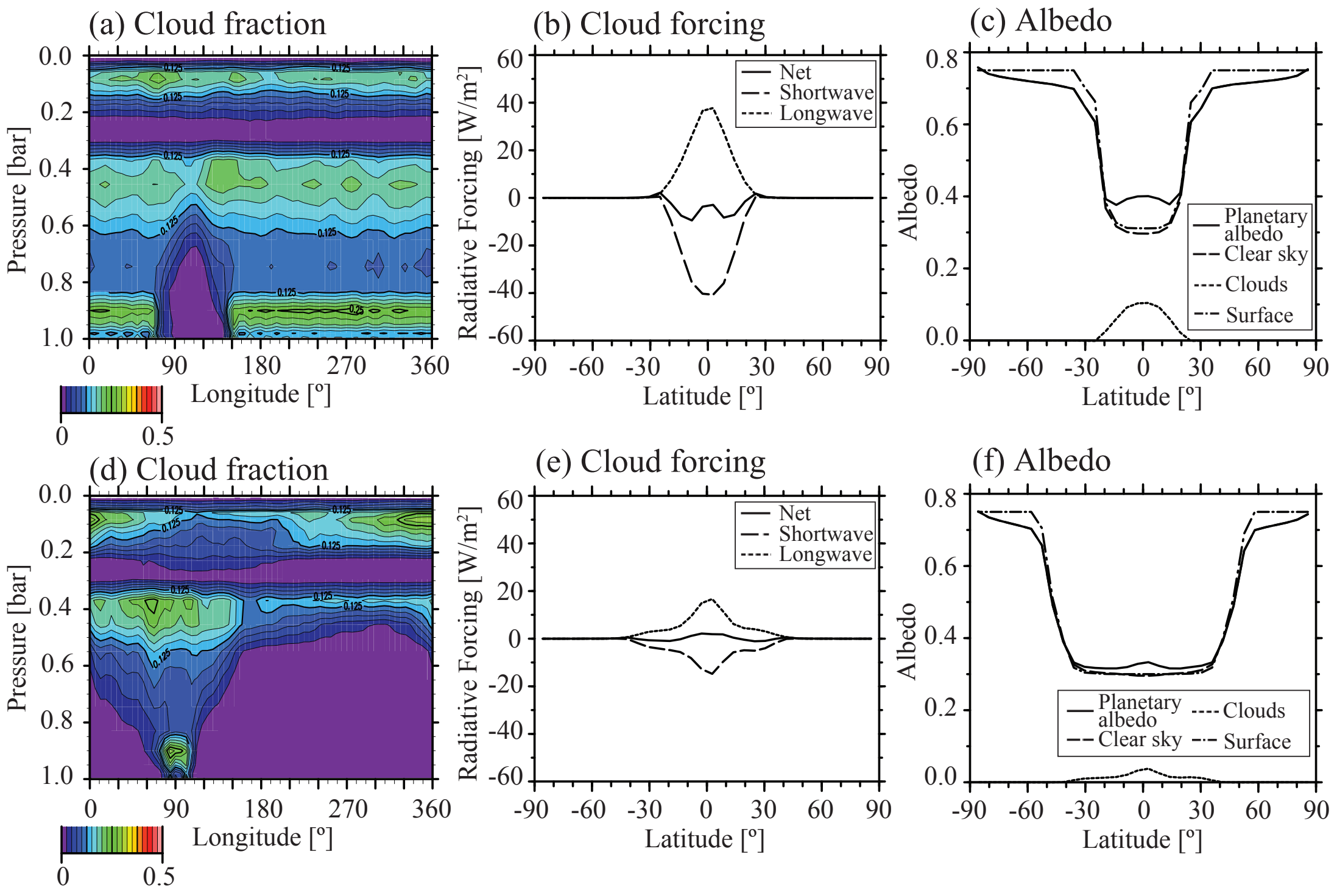

Figure 4. The meridionally averaged cloud fraction in the tropics (± 30º in latitude), the zonally averaged cloud radiative forcings and the albedos for cases of the meridionally concentrated surface water distributions with 90% $S_0$. Two different water distributions are shown in upper row for 52 grid cells and lower row for 6 grid cells of the water pool region.

Figure 5a-c show these variables at the complete freezing limit for a case of six grid cells in the water pool region. For this planet, the complete freezing limit is to be 80% $S_0$, which is somewhat different from the value (77% $S_0$) reported by Abe et al. (2011) and our result in Section 3.1. This difference is due to the behavior of the climate in the tropics. The tropical regions in Abe et al. (2011) and Section 3.1 are drier than for a meridional surface distribution. Kodama et al. (2018, 2019) concluded that the advection of water vapor to the tropical region is important for limiting planetary radiation because of the resulting optically thick atmosphere, causing a runaway greenhouse state. Moreover, we found that the advection of water vapor to the tropical region is significant for the complete freezing limit, since it leads to a higher cloud fraction and much snow in the water pool region. This drives positive feedback, similar to the ice-albedo feedback.

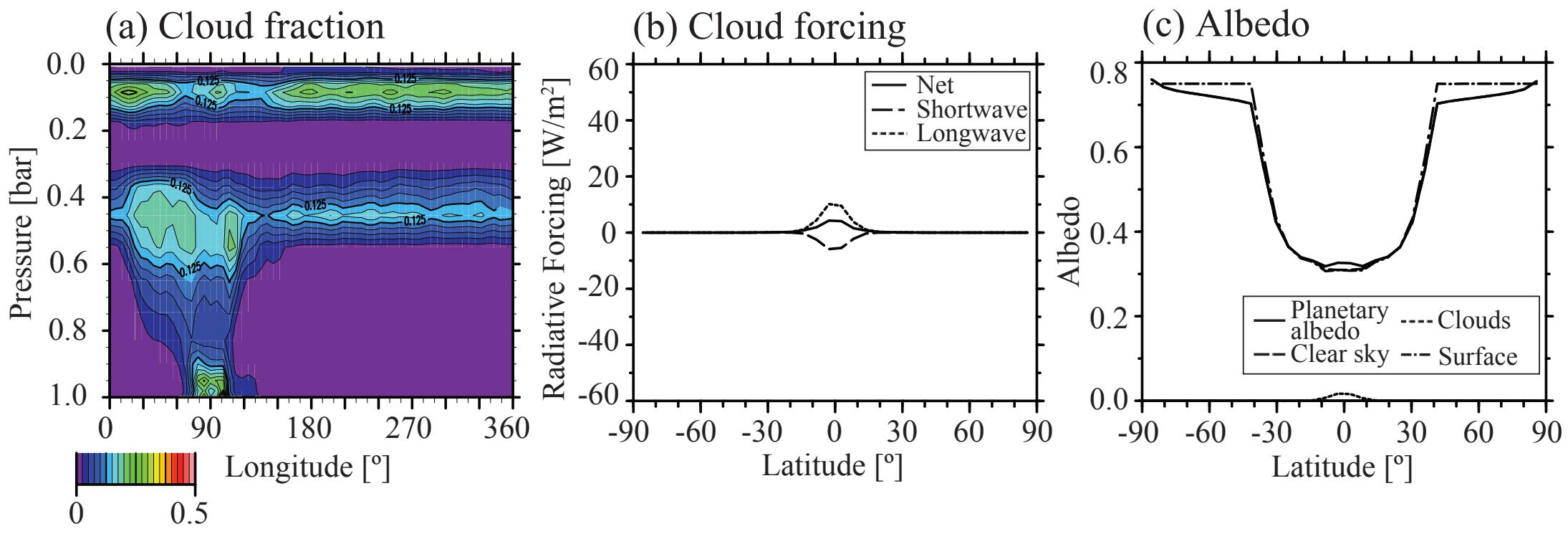

Figure 5. The meridionally averaged cloud fraction in tropics, the zonally averaged cloud radiative forcings and the albedos for a case of 6 grid cells of the water pool regions on the meridionally concentrated surface water distribution with 80% $S_0$.

Figure 6 shows the same variables as in Figs. 4 and 5 for the meridionally equally-dispersed case. In the case of 52 water pool grid cells, the complete freezing limit is 91% $S_0$, similar to that for an aqua planet. Although there are slight differences between the concentrated and dispersed cases with the same water pool area, the cloud contributions and snow distribution are similar (Fig. 4a-c and Fig. 6a-c). Both cases show that the distribution of snow reaches the tropic region and results in higher albedo. The explanation for the complete freezing limit is the same as in the case of the zonally uniform surface water distribution. For a case of six water pool grid cells in the meridionally equally-dispersed case, we show these variables at 91% $S_0$ in Figs. 6d-f. In Fig. 6f, there is a narrower snow distribution than for the 52 water pool grid cells in Figure 6c, causing a lower planetary albedo and a higher surface temperature to avoid lapsing into the complete freezing. We show these variables for six water pool grid cells at 81% $S_0$, which is the complete freezing limit for this case (Fig. 6g-i). Similar to other cases at the complete freezing limit, the distribution of snow reaches the tropics, making the planetary albedo higher.

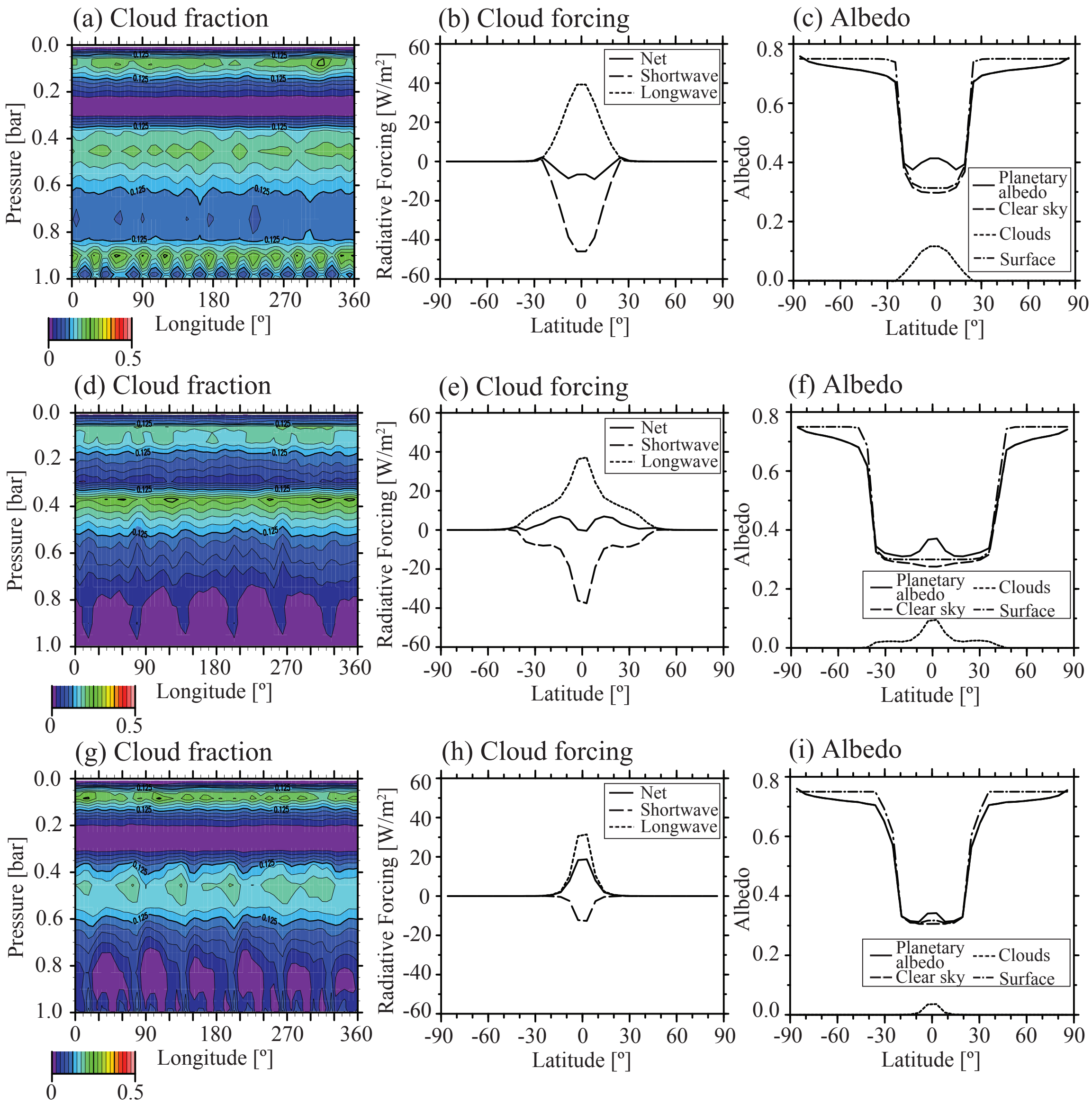

Figure 6. The meridionally averaged cloud fraction in tropics, the zonally averaged cloud radiative forcings and the albedos for the meridionally equally-dispersed case. Upper row shows a case for 52 grid cells of the water pool region at 91% $S_0$. Middle and bottom rows show cases for 6 grid cells of the water pool region at 91% and 81% $S_0$, respectively.

For the runaway greenhouse limit, Kodama et al. (2018, 2019) concluded that the distribution of water vapor in the tropics is important for governing the outgoing longwave radiation from the planetary atmosphere. On the other hand, for the complete freezing limit, the distribution of snow is a trigger for lapsing into the snowball state due to the transition from rainfall to snowfall in the tropics caused by decreasing temperatures.

### 3.3 Surface water distribution assuming planetary topographies

We also investigated the climate for surface water distributions with different amounts of water determined by particular planetary topographies in order to estimate the complete freezing limit. Figure 7 shows the same variables as in Fig. 3 for Earth-based cases, using the present ocean volume of Earth and 10% of that volume.

The complete freezing limit occurs at 90% $S_0$ for a planet with a surface water distribution equal to the Earth's present ocean volume of $1.37 \times 10^{18}$ $m^3$ (upper row in Fig. 7). Such a planet has a wide snow distribution and a high planetary albedo with a relatively large contribution of clouds to the planetary albedo, just as in the case of an aqua planet. On the other hand, for a planet with ~10% of Earth's present ocean volume ($10^{17}$ $m^3$), the distribution of snow is narrower and the cloud distribution is less than in the $1.37 \times 10^{18}$ $m^3$ case because of the surface water distribution (middle row in Fig. 7). As discussed above, such a planet has a warmer surface temperature and is resistant to complete freezing. However, at 84% $S_0$, this planet also lapses into the complete freezing state (lower row in Fig. 7). For Earth's topography, there are lower altitude regions around the tropics, which create the Pacific and Atlantic Oceans. Since water is stored to produce the water pool region in the tropics as the amount of water increases, the climate for this case is similar to the case for the meridionally uniform surface water distribution. The climates estimated for terrestrial topographies such as Mars and Venus are similar to Earth-based case since they exhibit nearly the same behavior in creating the water pool region.

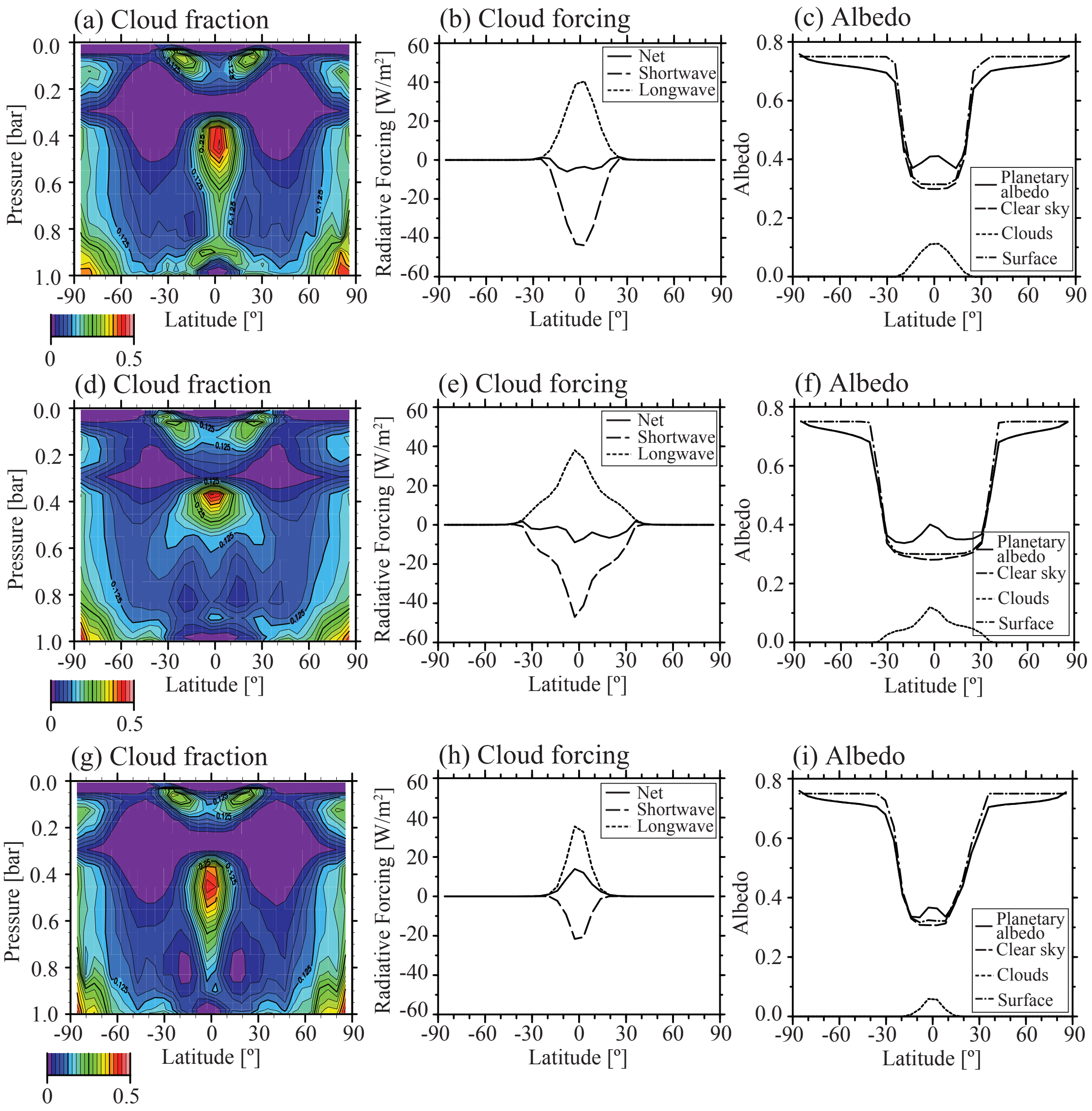

Figure 7. The zonally averaged cloud fraction, cloud radiative forcings and albedos for the surface water distribution using the planetary topographies for the Earth with different water amounts. Upper row shows a case for a planet with the present Earths' ocean volume ($1.37 \times 10^{18}$ m$^3$) of the amount of water at 90% $S_0$. Middle and bottom rows show cases for a planet with ~10% of the present Earths' ocean volume at 90% and 84% $S_0$, respectively.

## 4. Discussion

### 4.1 Stability of liquid water on the planetary surface

Figure 8 summarizes the complete freezing limit established in this study as a function of the initial land fraction, along with the runaway greenhouse limit from Kodama et al. (2018, 2019). These limits correspond to the condition of the existence of liquid water on the planetary surface with a reduced and increased insolation . The initial land fraction is the ratio of the initial dry area to the total surface area. A land fraction of zero indicates that the planetary surface is completely covered with liquid water.

We confirmed a range of variation for the insolation at the complete freezing limit, but found that the transition between climates for an aqua planet and a land planet is less clear. The insolation values at the complete freezing limit vary continuously from those for an aqua planet (~90% $S_0$) to those for a land planet (~77% $S_0$). Comparing the zonally uniform case to the meridionally uniform case, the results for the meridionally uniform case show consistently higher insolation at the complete freezing limit than was found for the zonally uniform case, as the meridionally uniform case has a wider water distribution in the tropics, causing ice-albedo feedback until the planet is in the complete freezing state.

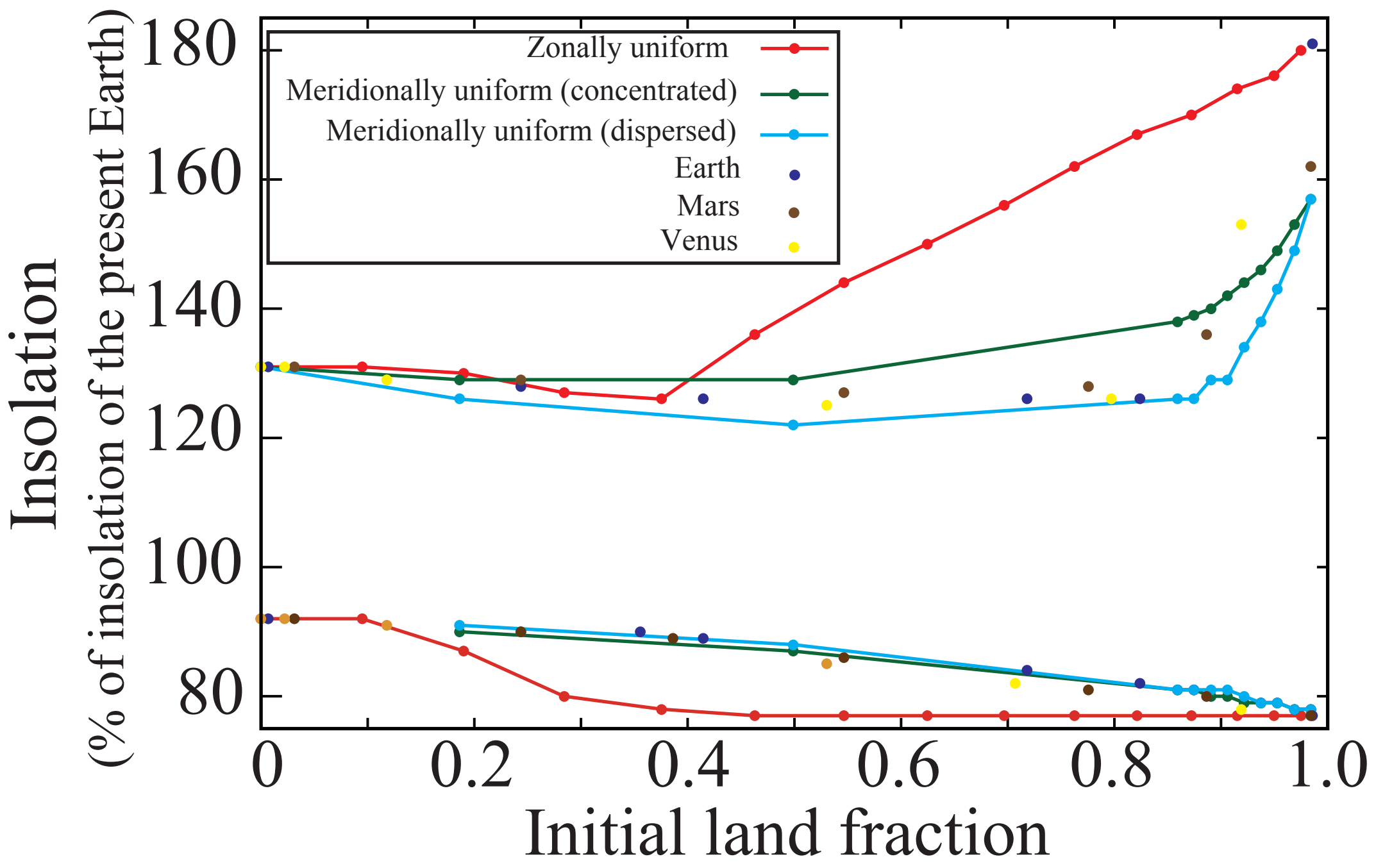

Figure 8. The insolations for the complete freezing limit and the runaway greenhouse limit. The solid lines show the insolation for both limits determined by the zonally and meridionally uniform surface water distributions. The insolation for the complete freezing limit determined by the topographies of the Earth, Mars and Venus, respectively, is located near the values for the meridionally uniformed cases.

We also estimated the complete freezing limit assuming a surface water distribution based on the topography of a terrestrial planet such as Earth, Mars, and Venus. For these cases, increases in the amount of water produce a water pool region in the low latitudes. The complete freezing limits here are between those in the zonally and meridionally uniform surface water distribution cases, with a value close to that for the case of the meridionally uniform surface water distribution, as the complete freezing limit is controlled by the cloud and snow distributions in the tropics, which are determined by the surface water distribution.

The runaway greenhouse limit is also strongly affected by the surface water distribution, and changes with a decrease in the wet surface area (Kodama et al., 2018, 2019). In comparing the zonally uniform case and the meridionally uniform case, we found that the runaway greenhouse limit for the latter is always lower than that for the former because of the distribution of water vapor, which leads to a limitation on planetary radiation. This means that the inner edge of the habitable zone has a width of insolation, depending on the surface water distribution. Kodama et al. (2019) also estimated the runaway greenhouse limit for surface water distributions assuming planetary topographies. These results assuming terrestrial planetary topographies are within a variation width of insolation at the inner edge of the habitable zone. For the runaway greenhouse limit, the boundary between climates for an aqua planet and a land planet is around an initial land fraction of 0.4, even though the transition is not clear for the complete freezing limit.

It is apparent that the water vapor distribution in the atmosphere is important for both limits for the existence of liquid water. It plays a significant role as a driver of ice (and cloud)-albedo feedback, which leads to a low surface temperature and heavy snow, ultimately resulting in the snowball state. On the other hand, water vapor plays a significant role as a limiter of planetary radiation, determining the threshold for the runaway greenhouse state. This would imply that both edges of the habitable zone are affected by the surface water distribution.

### 4.2 Amount of $CO_2$

In all of our simulations, we assumed an Earth-like atmosphere with 345 ppm of $CO_2$. As mentioned in Section 1, however, the amount of $CO_2$ is one of the key parameters in determining the width of the habitable zone, especially its outer edge. The concentration of $CO_2$ is still unclear in exoplanetary atmospheres because it depends on the amount of water that a planet has on its surface. From theoretical studies, a planet like Earth should have a carbonate-silicate cycle leading to stabilization of the $CO_2$ content in the atmosphere. A planet with a global ocean has a higher $CO_2$ concentration than a planet with both ocean and land (like Earth), since continental weathering is ineffective and an accumulation of $CO_2$ in the atmosphere will occur due to degassing from the interior (e.g., Tajika & Matsui, 1992). When a planet has sufficient water on its surface, high-pressure ice is formed on the seafloor. If this high-pressure ice layer is solid and completely prevents seafloor weathering, an accumulation of $CO_2$ in the atmosphere occurs, leading to a hot climate (Alibert, 2014; Kitzmann et al., 2015). However, if one considers the effects of sorbet flow on the liquid-solid coexistence region near mid-ocean ridges, such a planet would have a low surface temperature that leads to complete freezing even if the planet is located within the habitable zone (Nakayama et al., 2019). Furthermore, the degassing flux of $CO_2$ is important when discussing the evolution of a planetary climate. Depending on this flux, a planet within the habitable zone has the snowball limit cycle (e.g., Kadoya & Tajika, 2014; Kadoya & Tajika, 2015; Abbot 2016; Haqq-Misra et al., 2016; Kadoya & Tajika, 2019).

For a land planet, it is not clear whether there is a carbonate-silicate cycle since it is necessary to consider the recycling of materials between the planetary surface and the interior. The mantle on a land planet is drier than that on an aqua planet because the land planet has a higher land fraction. Assuming weaker tectonics, carbonates formed via the weathering process remain in nearly their original location. Thus, the carbonate-silicate cycle would not work effectively, even given that a land planet has more land than an aqua planet. In this case, an accumulation of $CO_2$ will occur, and the complete freezing limit will extend to a lower insolation. However, at a certain level of $CO_2$(approximately 8bar), the Rayleigh scattering, which cools the climate by reflecting the insolation, works efficiently. In this case, the complete freezing limit would increase due to the higher planetary albedo induced by the Rayleigh scattering. If a land planet has tectonics and a large degassing flux of $CO_2$, the concentration of $CO_2$ will decrease or remain at a constant level because of the higher surface temperature and wider land expanse than in the case of Earth, resulting in a higher value for the complete freezing limit. In any case, further studies are needed to investigate the effect of greenhouse gases on the climate of a land planet.

### 4.3 Water cycle

With respect to the inner edge of the habitable zone, the amount and distribution of water vapor in the tropics is significantly important in determining the threshold of the runaway greenhouse state caused by the imbalance of energy between the incoming stellar radiation and outgoing planetary radiation. For the complete freezing limit in this study, the surface water distribution is also important for determining the distribution of snow that causes a higher albedo in the tropics. Decreasing the insolation from the central star, the temperature in the atmosphere decreases and a transition from rainfall to snowfall occurs to create a wider snow cover area in the tropics. In this study, although we assumed that water and snow on the planetary surface do not move when rain and snow reach the surface, in reality, they move. If we consider the equatorward transport of surface water, the runaway greenhouse limit will be lower and the complete freezing limit will be higher than the limits reported in this study. For complete freezing, it would be important to consider movement of glaciers. We assumed a fixed maximum depth of snow in our series of simulations. If the snow cover is thick enough to move equatorward, then complete freezing would occur at a higher insolation than our results show, as the equatorward movement of glaciers would lead to a higher planetary albedo in the tropics.

### 4.4 Exoplanets

Recently, a few tens of terrestrial exoplanets within the classical habitable zone for an aqua planet have been detected. They are thought to be potential habitable exoplanets, meaning that such planets have a possibility that they would have liquid water on their surface. The climates of some of these exoplanets have been estimated using GCM, particularly in the case of Proxima Centauri b and the terrestrial planets in the TRAPPIST-1 system, (Turbet et al., 2016, 2018). Such potentially habitable planets are thought to have a synchronized rotation due to the tidal effect, as their host star is smaller than our Sun and thus the habitable zone is nearer to the host star. This leads to a permanent dayside and nightside on the planets. The trigger for the runaway greenhouse state to occur is an imbalance between the insolation from the central star and the outgoing planetary radiation due to the limit imposed on the outgoing planetary radiation, which strongly depends on the moisture content of the atmosphere. Yang et al. (2013) pointed out the importance of the presence of clouds on the dayside, especially around the substellar point, for maintaining a habitable climate around a low-mass star. Kopparapu et al. (2017) and Haqq-Misra et al. (2018) summarized the inner edge of the habitable zone considering the self-consistent orbital period for low-mass stars. From our results, a planet will lapse into a runaway greenhouse state when the tropics become moist. Therefore, the amount of water in the atmosphere around the substellar point is important for the inner edge of a tidally locked potentially habitable planet. When a planet has a wider dry region around the substellar point, meaning it is a land planet, it can radiate a greater amount of radiation than an aqua planet (i.e., the limit on planetary radiation is higher). Additionally, the nightside on such a planet will have a lower temperature, leading to the formation of an ice cap so long as the transportation of water vapor is suitably efficient. Depending on various physical mechanisms, which include ice flow and the geothermal flux, there is a possibility that such a planet will have long-lived liquid water on its surface, especially on the edge between the dayside and the nightside and/or at the bottom of an ice cap (Leconte et al. 2013b). We intend to address these effects in a subsequent paper.

In considering the evolution of potentially habitable exoplanets and their host star, the amount of water present on a planet is an important factor. According to theoretical studies on planet formation, the amount of water on terrestrial exoplanets varies over a wide range (e.g., Genda, 2016; Tian & Ida, 2015; Kimura & Ikoma, 2020). On the other hand, an aqua planet can lose liquid water in the moist greenhouse state before the onset of the runaway greenhouse state. Kodama et al. (2015) studied the evolutionary path of an aqua planet to a land planet due to water loss, finding that if the aqua planet loses a significant amount of water during the moist greenhouse state, it can still maintain liquid water on its surface as a land planet.

There are a number of important parameters in the formation of a planetary climate, including the planetary rotation rate, obliquity, and eccentricity. We suggest that atmospheric circulation and the distribution of water vapor in the atmosphere are significant factors in determining the habitability of terrestrial planets. Abe et al. (2005) investigated the effect of varying the planetary obliquity on climate and concluded that the boundary between an aqua planet and a land planet is determined by the width of the Hadley circulation, as it controls the regions where precipitation occurs. The width of the Hadley circulation is affected by several factors, such as the planetary rotation rate, temperature profile, and eddies (e.g., Satoh, 1994; Frierson et al., 2007; Held & Hou, 1980; Kaspi & Showman, 2015). If the planetary rotation is faster than assumed, the regions of the westerlies should become wider and the width of the Hadley circulation should become narrower. Thus, in this case, a planet would behave as a land planet, given the difficulty of spreading water vapor from the higher latitudes to tropical regions. Although the planetary obliquity and rotation rate have not yet been made clear from observations yet, future observations should provide sufficient information to establish these parameters (Kawahara, 2012; Kawahara, 2016; Nakagawa et al., 2020). Theoretical studies regarding the mapping of planetary surfaces have also been proposed (Fujii et al., 2010; Aizawa et al., 2020; Kawahara, 2020). In the next generation of studies on exoplanets, it should be possible to distinguish between an aqua planet and a land planet from observations.

## 5 Summary

In assessing the habitability of planets, most studies have focused on Earth-like planets with globally abundant liquid water on the planetary surface. Previous studies have shown that the surface water distribution is important for establishing the inner edge of the habitable zone, which corresponds to the runaway greenhouse limit (Abe et al., 2011; Kodama et al., 2018, 2019). However, although such prior studies have investigated the insolation for the runaway greenhouse limit for various surface water distributions, the relationship between the complete freezing limit and the surface water distribution has remained quantitatively unclear, and the surface water distribution should affect the complete freezing limit. In our investigation, we assumed the same three types of the surface water distributions as in Kodama et al. (2018) and Kodama et al. (2019)—a zonally uniform surface water distribution, a meridionally uniform surface water distribution, and a surface water distribution based on the topographies of the terrestrial planets in our solar system. Using a GCM, we systematically estimated the climates with reduced insolations for three types of the surface water distributions in order to explore the relationship between the insolation at the complete freezing limit and the surface water distribution.

In the case of zonally uniform surface water distributions, we recognized two climatic regimes for the complete freezing limit: an aqua planet climatic regime and a land planet climatic regime. This is similar to previous studies on the runaway greenhouse limit. We introduced the water flow limit, which is determined by the lowest latitude of the water pool region as a parameter. As described by Abe et al. (2011), dry planets have a drier atmosphere than aqua planets, leading to a low planetary albedo with less snow and fewer clouds. Just as the runaway greenhouse limit shows a wide variation depending on the surface water distribution, the complete freezing limit also has a range of values, from 90% $S_0$ for an aqua planet to 77% $S_0$ for a land planet, which is consistent with the results of Abe et al. (2011). The climate of a land planet is much warmer than that of an aqua planet due to the lower planetary albedo caused by the distribution of snow and clouds, causing the complete freezing limit to be lower.

In our treatment of meridionally uniform surface water distributions, we assumed two cases: a concentrated case and an equally-dispersed case. As in the zonally uniform case, complete freezing occurs when the distribution of snow reaches the tropics, producing a higher albedo. Because meridionally uniform surface water distributions have water pool regions in the tropics, these cases tend to trigger an ice-albedo feedback that causes a planet to lapse into a complete freezing state at a higher insolation than in the zonally uniform surface water distribution cases. For the complete freezing limit, the distribution of snow is trigger for lapsing into the complete freezing state due to the transition from rainfall to snowfall in the tropics caused by the deceasing temperatures. Thus, the complete freezing limits for the meridionally uniform cases are consistently higher than in the zonally uniform cases. We also estimated the complete freezing limit for the surface water distribution with different amounts of water based on the planetary topographies of Earth, Mars, and Venus. The complete freezing limits for these cases are close to those for the meridionally uniform cases because of the surface water distribution.

We confirmed the complete freezing limit with different surface water distributions and showed that the surface water distribution is significantly important for determining both limits of stability for liquid water on the planetary surface with reduced insolations. The amount of water present on a planet is difficult to estimate from its expected bulk composition based on the relationship between planetary mass and radius. However, differences in the amount of water present on a planet can create significant climatic differences. If future observations allow us to distinguish these differences, it will be possible to establish more clearly the constraints that the amount and distribution of surface water impose on the surface environment.

## Acknowledgments

We thank the editor and two reviewers for their constructive comments and suggestions. This work was supported by MEXT KAKENHI Grants JP21K13975 and the Astrobiology Center of National Institutes of Natural Sciences (NINS) (Grant Number AB031014). H. Genda acknowledges the financial support of MEXT KAKENHI grant No. JP17H06457. A. Abe-Ouchi was supported by MEXT KAKENHI grant No. JP17H06104. This project has also received funding from the European Research Council (ERC) under the European Union's Horizon 2020 research and innovation programme (Grant Agreement 679030/WHIPLASH). The GCM used in this study is CCSR/NIES AGCM 5.4g which has been developed for the Earth's climate by the Center for Climate System Research, the University of Tokyo, and the National Institute for Environmental Research. We used Gtool-3-dc5 which is developed by GFD-DENNOU Club for analysis (https://www.gfd-dennou.org/library/gtool/index.htm.en). Data sets to create typical figures in this study are available on website: Kodama, Takanori. (2021). Datasets of "The Onset of a Globally Ice-covered State for a Land Planet" [Data set]. Zenodo. (https://doi.org/10.5281/zenodo.5644572). For more detail, please see the Supporting Information.